\documentclass[epj]{svjour}
\usepackage{graphics}
\begin{document}
\title{Birth and long-time stabilization of out-of-equilibrium
coherent structures}
\author{
Julien Barr{\'e}\inst{1,2}, Freddy Bouchet\inst{2,4}, Thierry
Dauxois\inst{1}\thanks{Thierry.Dauxois@ens-lyon.fr} \and Stefano
Ruffo\inst{2,3}
}                     
\offprints{Thierry Dauxois}          
\institute{Laboratoire de Physique, UMR-CNRS 5672, ENS Lyon, 46
All\'{e}e d'Italie, 69364 Lyon C\'{e}dex 07, France \and
Dipartimento di Energetica "S. Stecco", Universit{\`a} di Firenze,
via S. Marta, 3, I-50139 Firenze, Italy \and INFM and INFN, Firenze,
Italy
\and UMR 5582, Institut Fourier 74, 38402 Saint Martin d'H{\`e}res C{\'e}dex, France}
\date{Received: date / Revised version: \today}
%
\abstract{We study an analytically tractable model with {\it long-range
interactions} for which an out-of-equilibrium very long-lived
coherent structure spontaneously appears. The dynamics of this model is
indeed very peculiar: a {\it bicluster} forms at low energy and is stable
for very long time, contrary to statistical mechanics predictions. We first
explain the onset of the structure, by approximating the short time
dynamics with a forced Burgers equation.  The emergence of
the bicluster is the signature of the shock waves present in
the associated hydrodynamical equations. The striking quantitative
agreement with the dynamics of the particles fully confirms this
procedure. We then show that a very fast timescale
can be singled out from a slower motion. This enables us to use an
adiabatic approximation to derive an effective Hamiltonian
that describes very well the long time dynamics. We then get an 
explanation of the very long time stability of the bicluster:
this out-of-equilibrium state corresponds to a statistical equilibrium 
of an effective mean-field dynamics.
Classical rotators. XY model.
} 
\authorrunning{Barr{\'e}, Bouchet, Dauxois, and Ruffo}
\titlerunning{Out-of-equilibrium states as statistical equilibria}
\maketitle

\section{Introduction}
\label{Introduction}
The Hamiltonian Mean Field model (HMF) has attracted much attention in
the recent years as a toy model to study the dynamics of systems with
{\it long-range} interactions, and its relation to thermodynamics
\cite{Antoni,AnteneodoTsallis,LatoraRapisardaRuffo}. The HMF model
describes an assembly of $N$ fully coupled rotators, whose Hamiltonian is:
\begin{eqnarray}
H & = & \sum_{i=1}^{N}\frac{p_i^2}{2}+\frac{c}{2N}\sum_{i,j=1}^{N}\cos(\theta_i-\theta_j)\quad,
\label{hamiltonian}
\end{eqnarray}
where $\theta_i$ is the angle of the $i$-th planar rotator with a fixed
axis. As the interaction only depends on the angles of the
rotators, this model can alternatively be viewed as representing
particles that move on a circle, whose positions are given by the
$\theta_i$ and interact via an infinite-range force. If $c$ is
negative, the interaction among rotators is ferromagnetic,
corresponding to an attractive interaction among particles. When
$c$ is positive, the interaction among rotators is
antiferromagnetic and repulsive in the particle interpretation. In
this article, we will focus our study on this latter case, and put
$c=1$.

\begin{figure}
\resizebox{0.6\textwidth}{!}{\includegraphics{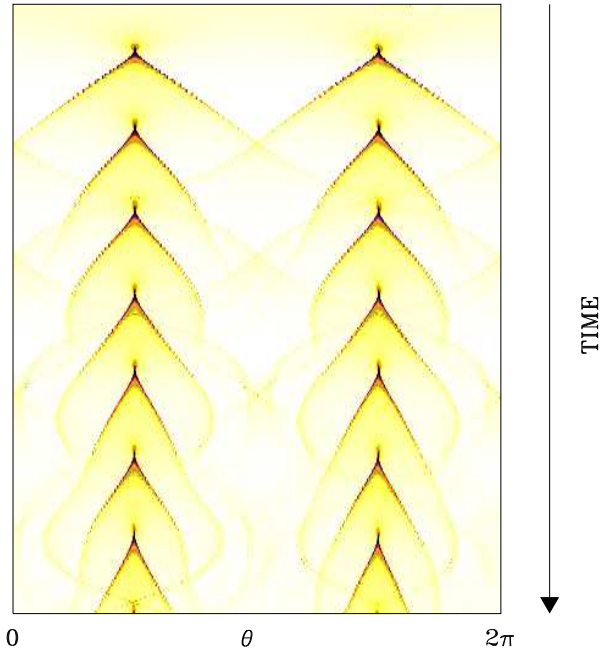}}
\caption{Bicluster formation : short-time evolution of the particle density
in grey scale: the darker the grey, the higher the density. Starting from
an initial condition with all the particles evenly distributed on the circle,
one observes a very rapid concentration of particles, followed by the
quasi periodic appearance of ``chevrons'', that shrink as time increases.}
\label{chevrons1}
\end{figure}

As first noticed in Ref.~\cite{Antoni}, this model has a very
interesting dynamical behavior. In contrast to the statistical
mechanics predictions, a {\it bicluster} forms at low energy (see
Fig.~\ref{chevrons1}), for a special, but wide class of initial
conditions: it appears as soon as the initial velocity dispersion of
the particles is small, for any initial spatial distribution.  This
clustering phenomenon and its unexpected dependence on initial
conditions were precisely studied in Ref.~\cite{drh}, but remained
essentially unexplained. Several details of the dynamics were
numerically studied and it has been in particular emphasized that the
energy temperature relation is modified: although still linear, the
slope in the molecular dynamics simulations is different from the
theoretical prediction. No sign in the numerics was found of the decay
to the constant density profile predicted by the canonical ensemble.
On the contrary, recently performed simulations~\cite{FFR} with a
smaller number of particles have shown a long-time degradation of the
bicluster, suggesting its transient non-equilibrium nature. The
question of the time asymptotic stability of the bicluster, in the
limit of an infinite number of particles, remains however open and
will be discussed in the conclusions.

We will consider in this work two different theoretical
approaches that lead to an analytical explanation of the
clustering process and of the long time stability
of the structure. In section~\ref{Hydrodynamicaldescription}, in
order to explain the short time bicluster formation, we first
propose a hydrodynamical description that links
Hamiltonian~(\ref{hamiltonian}) to a forced Burgers equation
(these results were shortly presented in a previous
brief note~\cite{BDR}).
This striking phenomenon is then described in
Lagrangian coordinates, giving rise to intersections of
trajectories and formation of high (or
even infinite) particle densities. We then solve in
section~\ref{characteristic} this equation, using
extensively the method of characteristics. Beyond the
theoretical interest in such structures for Hamiltonian systems, we
will show some connections of this problem with other subjects;
namely, active transport in hydrodynamics, and the
formation of caustics in the flow of the associated Burgers equation.

Section~\ref{lagrangianeff} presents the second approach, which is
particularly powerful for explaining the long-time evolution of the
bicluster. The dynamics involves two well separated time-scales. Using
a variational approach, we will show how to construct an effective
Hamiltonian where all the fast oscillations are averaged out. Not only
this effective dynamics accurately represents the one of the original
Hamiltonian but, in addition, the complete statistical thermodynamics
can be easily derived in the microcanonical and canonical ensembles.
It predicts the existence of the bicluster as an equilibrium state,
with the correct density profile and the microcanonical temperature
energy relation found in the numerical experiments of Ref.~\cite{drh}.
The long lifetime of these out-of-equili\-brium states can therefore be
interpreted by the fact that they appear as equilibrium states of an
effective Hamiltonian representing the long-time motion.
Section~\ref{Sec-Conclusions} is devoted to some conclusions and
perspectives of future developments.

\section{Hydrodynamical description at zero temperature}
\label{Hydrodynamicaldescription}
\subsection{Introduction}
To describe the initial  clustering process, we consider the
associated Vlasov equation, which can be rigorously derived in the
thermodynamic limit
$N\longrightarrow \infty$~\cite{Kinlimit,BraunHepp}.  Denoting by $f(\theta,p,t)$
the
one particle
distribution function, we have:
\begin{equation}
\frac{\partial f}{\partial t}+p \frac{\partial f}{\partial \theta}
-\left[\frac{1}{2\pi} \int_{-\infty}^{+\infty}du \int_{0}^{2\pi}
d\alpha \ f(\alpha,u,t) \sin(\theta-\alpha)\right]
 \frac{\partial f}{\partial p} = 0\label{Vlasov}
\end{equation}
Let us define  as follows, a density
\begin{equation} \rho(\theta,t) =
\int_{-\infty}^{+\infty}f(\theta,p,t)\;dp\label{defrho}
\end{equation} and a velocity
field
\begin{equation}
\rho(\theta,t)v(\theta,t) =
\int_{-\infty}^{+\infty}pf(\theta,p,t)\;dp\quad.\label{defvit}\end{equation}
As the numerical simulations reported in Ref.~\cite{drh} have shown that the
bicluster appears when the velocity dispersion is small, 
we will consider here the zero temperature approximation, i.e. we
neglect the velocity dispersion and consider the ansatz
$f(\theta,p,t)=\rho(\theta,t) \delta(p-v(\theta,t))$, where $\delta$ refers 
to the Dirac function. The Vlasov
equation~(\ref{Vlasov}) can consequently be simplified to
\begin{equation}
\frac{\partial f}{\partial t}+p \frac{\partial f}{\partial \theta}
-F(\theta,t)\; \frac{\partial f}{\partial p} = 0\label{Vlasovsimpl}
\end{equation}
where one notes
\begin{equation}\label{defrhobar}
F(\theta,t)=\frac{1}{2\pi}  \int_{0}^{2\pi} d\alpha \
\rho(\alpha,t) \sin(\theta-\alpha)\quad.
\end{equation}

 The sum of
the time derivative of Eq.~(\ref{defrho}) and the $\theta$-derivative
 of Eq.~(\ref{defvit}), leads to
\begin{eqnarray}\label{tyi}
  \frac{\partial \rho}{\partial t} + \frac{\partial (\rho v)}{\partial
\theta}
  &=&
  \int_{-\infty}^{+\infty}\left(\frac{\partial f}{\partial t}+p \frac{\partial f}{\partial \theta}\right)dp\nonumber\\
  &=& \int_{-\infty}^{+\infty}dp\ F(\theta,t)
 \frac{\partial f}{\partial p}=0\quad.
\end{eqnarray}
We obtain therefore the equation
which accounts for mass conservation
\begin{equation}
  \label{conser}
  \frac{\partial \rho}{\partial t} + \frac{\partial (\rho v)}{\partial
\theta} = 0\quad.
\end{equation}

Using the time derivative of Eq.~(\ref{defvit}) and Eq.~(\ref{conser}),
we obtain
\begin{equation}
   \frac{\partial\left( \rho v\right)}{\partial t} =\rho\left(\frac{\partial v}{\partial t} + v\frac{\partial v}{\partial \theta}\right) -\frac{\partial (\rho v^2)}{\partial
\theta} \quad.\label{tyuioorih}
\end{equation}
However, the time derivative of the left-hand-side gives also
\begin{eqnarray}
   \frac{\partial\left( \rho v\right)}{\partial t}
   &=&\int_{-\infty}^{+\infty}p\frac{\partial f}{\partial
     t}\;dp\nonumber\\
&=&-\int_{-\infty}^{+\infty}p^2\frac{\partial f}{\partial
     \theta}\;dp+F(\theta,t)\int_{-\infty}^{+\infty}p\frac{\partial f}{\partial
     p}\;dp\nonumber\\
&=&-v^2\frac{\partial \rho}{\partial
     \theta}-\rho\frac{\partial v}{\partial
     \theta}2v +F(\theta,t){\rho}(\theta,t)\quad.\label{tyuio}
\end{eqnarray}
Equations~(\ref{tyuioorih}) and Eq.~(\ref{tyuio}) lead finally to
 the following Euler equation without pressure term:
\begin{eqnarray}
&&\frac{\partial v}{\partial t} + v\frac{\partial v}{\partial \theta}
=\frac{1}{2\pi} \int_{0}^{2\pi} d\alpha \
\rho(\alpha,t)\sin(\theta-\alpha) \label{euler}
\end{eqnarray}
Equations~(\ref{conser}) and~(\ref{euler}) show that the dynamics of
the model at low temperature can be therefore mapped onto an active
scalar advection problem.

\subsection{Linear analysis}
\label{linear}
Linearizing equations (\ref{conser}) and (\ref{euler}), by assuming
small velocities ($v\ll1$) and almost uniform density ($\rho=1+\rho_1(\theta,t)$ with $\rho_1\ll
1$), we are left with:
\begin{eqnarray}
&&\frac{\partial \rho_1}{\partial t}  +  \frac{\partial v}{\partial \theta} =  0 \\
&&\frac{\partial v}{\partial t} =
\frac{1}{2\pi}\int_0^{2\pi}\rho_1(\alpha,t)\sin(\theta-\alpha)
\end{eqnarray}
a system which is easily solved by a spatial Fourier series
development. Assuming a zero initial velocity field, we get
\begin{eqnarray}
&& \rho_1(\theta,t) =  \sqrt2\ \ v_m\cos\theta\cos\omega t \label{linearanalya}\\
&& v(\theta,t) = v_m\sin\theta\sin\omega t~,\label{linearanalyb}
\end{eqnarray}
where the time-scale of the oscillation is given by the inverse of a sort
of ``plasma frequency" $\omega=\sqrt 2/2$, which describes, as in the Poisson-Boltzmann
case, the instability of the uniform density state~\cite{Nicholson}. In the
attractive case ($c=-1$ in Hamiltonian (\ref{hamiltonian})), this time
scale would control the depart from the initial uniform density towards
the formation of the density profile of the single cluster that appears at low
temperature~\cite{Ina1,Ina2}, in analogy with Jeans instability in
gravitational systems~\cite{Binney}.

This linear analysis is, however, not sufficient to explain the formation of
the bicluster and we have to carry out a non linear analysis.

\subsection{Non linear analysis}

\label{nonlinear}

This analysis relies on the existence of two time-scales in the
system: the first one is intrinsic and corresponds to the inverse of
the ``plasma frequency" $\omega$; the second one is connected to the
energy per particle $e$. When the energy is sufficiently small,
these time-scales are very different, and it becomes possible to
use averaging methods. The presence of two well separated
time-scales also explains, in some sense, the appearance of
the bicluster in the low energy limit.

We introduce therefore a long time-scale $\tau=\varepsilon t$,
with $\varepsilon= v_m/\sqrt 2=\sqrt{2e}$, and we look for
solutions of the following form:
\begin{eqnarray}
\label{vitesse} v(\theta,t) & = & v_m\sin\theta\sin\omega t
+ \varepsilon\ u(\theta,\tau)\quad .
\end{eqnarray}
In order to evaluate the force on the r.h.s. of
Eq.~(\ref{euler}) in the nonlinear regime, we will use
expressions~(\ref{linearanalya}) and~(\ref{linearanalyb}), given
by the linear analysis.  Indeed, due to the special form of the
Hamiltonian, the force depends only on the first Fourier component
of the density. Hence, our hypothesis amounts to assume that the
sinuso\"{\i}dal behavior of the density, found in the linear regime, holds
the same in the non-linear regime: the results presented below
confirm the validity of this assumption.

In analogy with studies of the wave-particle interaction in plasma
physics~\cite{plasma,firpo}, our system may be seen as a bulk of
particles interacting with waves that are sustained by the bulk
itself. Here, the wave is created by the small density and
velocity oscillations  already present in the linear analysis. In
the low energy regime, the phase velocity of the wave with the
plasma frequency $\omega$ is much higher than the velocities of
the bulk particles, causing a very small wave-particle interaction
strength: the wave has therefore a very long life-time. This also
explains the quality of the linear approximation for the force
that we have used in the Euler equation~(\ref{euler}).

Introducing expression~(\ref{vitesse}) in Eq.~(\ref{euler}),
terms to first order in $\varepsilon$ disappear by construction.
Order $\varepsilon^2 $ terms give:
\begin{eqnarray}
\frac{\partial u}{\partial \tau} + u\frac{\partial u}{\partial
\theta}&+&2\sin\theta \cos\theta \sin^2\omega t\nonumber\\
&+&\left(\sin\theta \frac{\partial u}{\partial \theta}+u\cos\theta
\right) \sqrt{2} \sin\omega t  = 0
\end{eqnarray}
Averaging over the short time scale~$t$, we obtain:
\begin{eqnarray}
\label{burgers} \frac{\partial u}{\partial \tau} +u\frac{\partial u}
{\partial \theta} & = & -\frac{1}{2}\sin 2\theta
\end{eqnarray}
which is a spatially forced Burgers equation without viscosity,
describing the motion of fluid particles in the potential
$V(\theta)=1/4\, \cos2\theta$.

The Burgers equation has been studied in many different contexts by
mathematicians interested in a pressureless description of fluid motion,
and by physicists in the context of ballistic
aggregation models~\cite{young}. In
particular, the forced Burgers equation has been very carefully
studied from a mathematical viewpoint~\cite{moserjauslin}.
Cosmologists have proposed a very interesting adhesion model
that uses Burgers-like equations to explain structure formation
in the universe~\cite{youngIHP}.
Moreover, this equation may be also viewed as a nice model of
active transport of the vorticity field in fluid mechanics~\cite{Castillo}.

\begin{figure}
\resizebox{0.5\textwidth}{!}{  \includegraphics{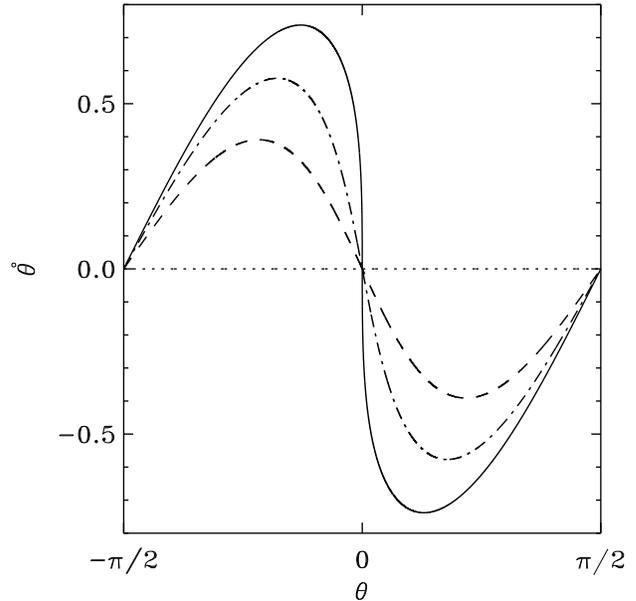}}
\caption{Shock Dynamics. Phase space portrait of $N=10^4$ particles, which
  at $\tau=\sqrt{2e}t=0$ are uniformly distributed in space (dotted line)
  with a small sinusoidal velocity profile, not visible in the
  figure. The resulting energy is $e=6.7\ 10^{-4}$.  Velocity
  profiles $u(\theta)$ are then shown at  $\tau=\pi/4$ (dashed
  line), $\tau=3\pi/8$ (dashed-dotted line) and $\tau=\tau_s=\pi/2$, the first
  shock time (solid line). Only half the space is shown since the
  curves are $\pi$-periodic.} \label{schock}
\end{figure}

\begin{figure}
\resizebox{0.5\textwidth}{!}{  \includegraphics{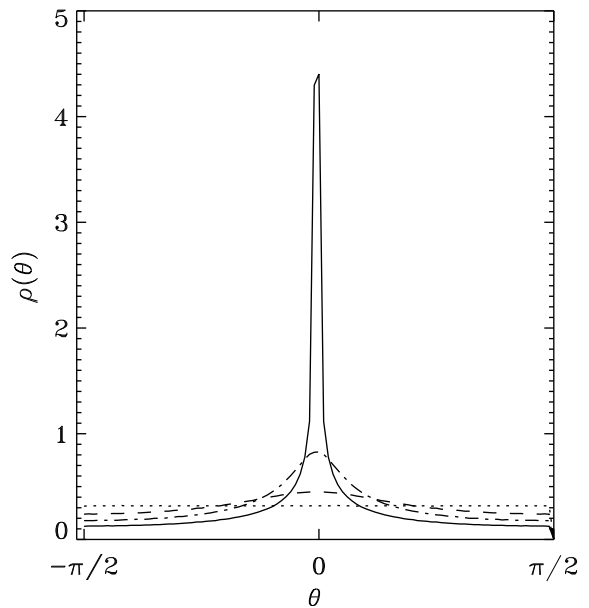}}
\caption{Particle density $\rho(\theta)$ at  $\tau=0$
(dotted line),  $\tau=\pi/4$ (dashed line), $\tau=3\pi/8$
(dashed-dotted line) and $\tau=\tau_s=\pi/2$ (solid line).
The initial condition is the same as in Fig.~\ref{schock}.
We clearly see the appearance of one of the two density peaks of the
bicluster.} \label{density}
\end{figure}

\begin{figure}
\resizebox{0.5\textwidth}{!}{  \includegraphics{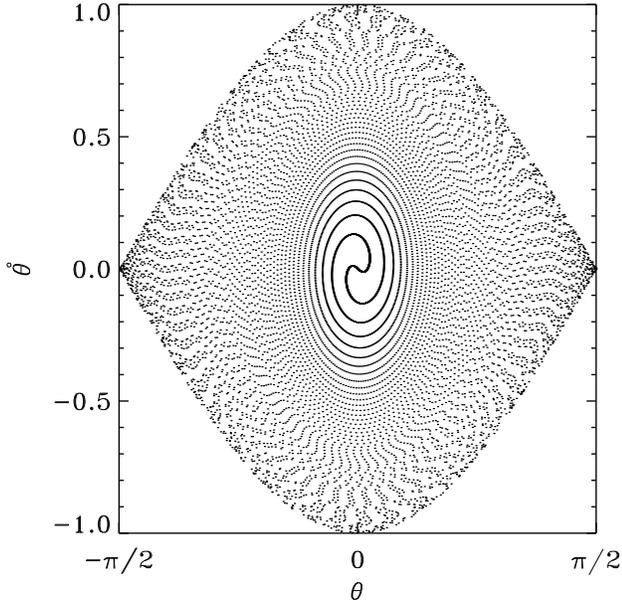}}
\caption{Spiral Dynamics. Phase space portrait of $N=10^4$
particles at $\tau=500$. The initial condition is the same
as in Fig.~\ref{schock}.}\label{spirale}
\end{figure}

A well known property of the Burgers equation without viscosity, is
that the solution becomes multi-stream after a {\em finite} time: the
appearance of shocks (see Fig.~\ref{schock}) in the velocity profile
$u(\theta)$ corresponds to the creation of singularities in the density
profile (see Fig.~\ref{density}). This is a consequence of the two
main assumptions: the medium is supposed to be continuous and the
temperature  is set at zero. Weakening either
assumption would have smoothed out the singularities.  In the original {\em
  discrete} Hamiltonian model, particles can cross and, after some
time, those that travel faster can be catched by the slower ones that
lie downstream, creating the {\it spiral dynamics} exemplified in
Fig.~(\ref{spirale}). The particle density has a peak at the center of
the spiral (Fig.~\ref{density}), and diverges at the peak position 
in the $N\to\infty$ limit at the shock time~$\tau_s$.  The
presence of a shock at finite time in the forced Burgers equation does
not prevent the hydrodynamical description from remaining valid at
longer times, when, moreover, several repeated shocks are observed at
regular time intervals. We will indeed show that it can adequately
describe even the long-time evolution.  In addition, let us remark
that it's the double well shape of the potential that forces the
Burgers equation, which is responsible for the formation of {\em two}
clusters.  Starting from the initially uniform state, the particles
will move around the bottom of the both wells and the majority of them
happen to be there at the same time at the shock time, creating
therefore two coherent structures that after several oscillations form
the bicluster at long-time.

\section{Solution of the spatially forced Burgers equation by
the method of characteristics} \label{characteristic}

The method of characteristics is the appropriate mathematical tool to
study more quantitatively the forced Burgers equation (\ref{burgers}),
but it has also a direct physical meaning: the characteristics
correspond to the Lagrangian trajectories of the Euler equation,
and are therefore
good approximations for the particle trajectories of the finite~$N$
Hamiltonian system (\ref{hamiltonian}) when $N\gg1$.

\begin{figure}
\resizebox{0.5\textwidth}{!}{
\includegraphics{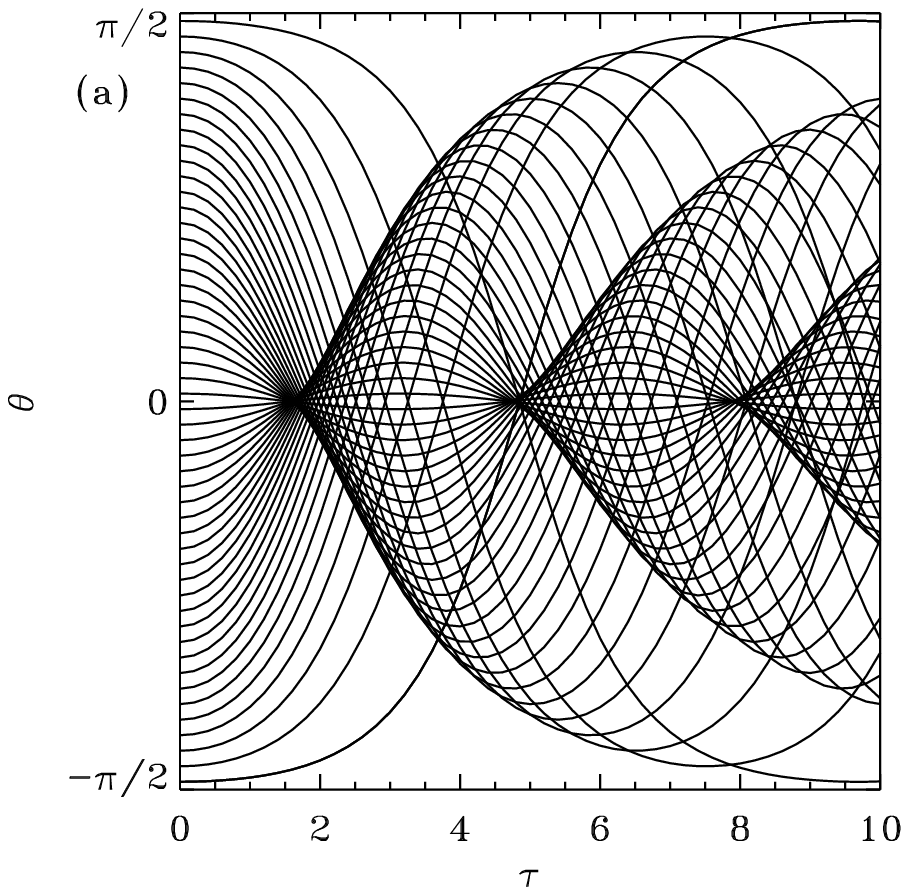}} \resizebox{0.5\textwidth}{!}{
\includegraphics{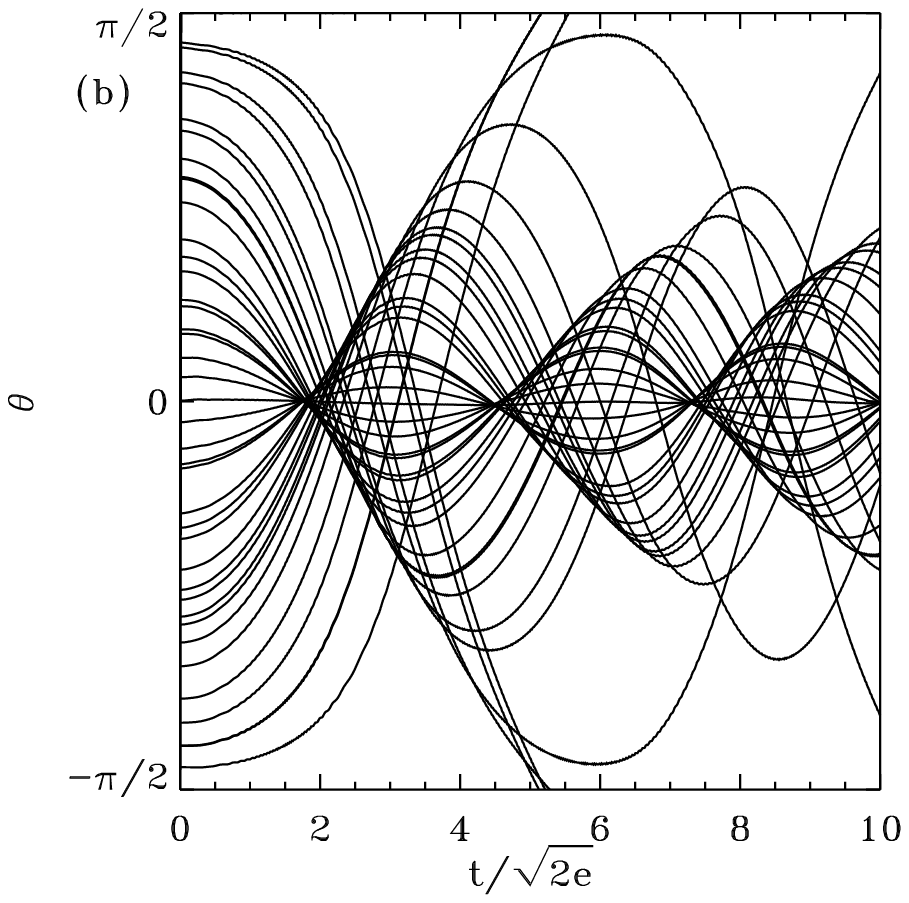}}
\caption{Panel (a) presents the characteristics
of equation (\ref{burgers}) whereas panel (b) presents
the trajectories of the particles of
the Hamiltonian  (\ref{hamiltonian}) with the same
initial condition as in Fig.~\ref{schock}. One can see the three first
appearances of the ``chevrons''. Two phenomena are not captured by
the characteristics: the small oscillations of the real
trajectories, which are averaged out, and the presence of
untrapped particles, close to the saddle-points of the effective
$\cos2\theta$ potential.} \label{carac}
\end{figure}

In the case of unidirectional nonlinear wave motions, the method is
standard and proceeds as indicated, for example, in
Ref.~\cite{Zauderer}.
We obtain in this case the following pendulum equation for
the characteristics $\theta(\tau)$:
\begin{eqnarray}
\frac{d^2\theta}{d\tau^2} + \frac{1}{2} \sin 2\theta & = & 0~,
\label{burgerslag}
\end{eqnarray}
with the initial conditions $\theta(0)=\theta_0$ and $\dot\theta(0)=0$.
The solution of the above equation can be easily
derived~\cite{lawden} and reads:
\begin{equation}\label{solelliptic}
  \theta=\arcsin\left[ \sin\theta_0\
  \mbox{sn}\left(\tau+K_0,\sin\theta_0\right)\right]
\end{equation}
where $K_0=K(\sin \theta_0)$, $K$ is the complete elliptic integral of
the 1st kind and sn, the elliptic sine function.
Fig.~(\ref{carac}a) shows the characteristics, i.e. the solutions
of Eq.~(\ref{burgerslag}) corresponding to different initial
conditions $\theta_0$ evenly distributed in $[0,2\pi]$ at $t=0$
and with zero initial velocity. One clearly sees their oscillatory
behavior inside one of the two wells of the effective potential
$\cos2\theta$, located at $\theta=0$, with a period $4K_0$ which strongly depends on the initial position.
Fig.~(\ref{carac}a) emphasizes also that the characteristics cross
themselves, and the associated {\it caustics} are responsible for
the enhancement of the particle density field that forms the
``chevrons". The time of the first divergence in the density $\tau_s$
is the time of the first characteristic crossing and also the time
of the first shock in the velocity profile. The particles
performing a quasi-harmonic oscillatory motion in the bottom of
the well are the ones with shorter periods. The shortest period is
obtained in the harmonic limit $4 K(0)=2\pi$, leading to
$\tau_s=\pi/2$ for the time of appearance of the first shock. In
the original units, it reads
\begin{equation}
t_s=\frac{\tau_s}{\varepsilon}=\frac{\pi}{\sqrt{8e}}\quad .
\end{equation}

This shock results in a singularity in Eulerian space at the
corresponding time.  However this singularity disappears immediately
after forming and two singularities of another kind arise in its
place, which are the boundaries of the three streams region. The
chevron is their manifestation in the density profile.  The shock time
singularity is called $A_3$ in Arnold's
classification~\cite{arnold,Zeldovich} while the ``chevrons''
singularity is of $A_2$ type. The latter is the boundary of structures
and can exist at any moment of time, whereas the former exist only
when a chevron originates.  The recurrence time for the appearance of
chevrons is $t_n=t_s*(2n-1)$ and the comparison with the numerics shows a
very good agreement.

Figs.~(\ref{carac}a) and ~(\ref{carac}b) show that the agreement is
really good not only at the qualitative but also at the quantitative
level. Only two features are missed by the Lagrangian description. The
fast oscillations of very small amplitude, already hardly visible in
Fig.~(\ref{carac}b) is of course totally absent in the characteristics
because of the averaging procedure used to get Eq.~(\ref{burgers}).
One can also easily show that the amplitude of this fast oscillation
is rapidly decreasing with the increase of the number of particles and
is vanishingly small for $N\to\infty$.  The second point is the presence
of untrapped particles in Fig.~(\ref{carac}b), which is a direct
consequence of the oscillations of the height of the potential
barriers. This effect is important only for highly energetic particles
whose trajectories in phase space are closed to the separatrix of the
Kelvin's cat eye of Fig.~(\ref{spirale}). However, as we will show in
the next section, these particles do not take part to the creation of
caustics (i.e. of the chevrons of Fig.~\ref{chevrons1}) which are
generated only by the particles close to the bottom of the wells.

\subsection{The chevrons as caustics of the characteristics}
We will now explain the shape of the "chevrons" (Fig.~\ref{chevrons1}), which
correspond to zones of infinite density i.e.  to the envelops of
the characteristics, the so-called caustics. The family of the
characteristics is defined by
\begin{equation}F(\tau,\theta,\theta_0)=\sin
\theta(\tau,\theta_0)-\sin(\theta)=0\end{equation}  in the plane
$(\theta,\tau)$ with the parameter $\theta_0$. The envelope of this
family would then be defined by the two following equations
\begin{eqnarray}\label{defenvelop}
F(\tau,\theta,\theta_0)&=&0\\
\frac{\partial F}{\partial \theta_0}(\tau,\theta,\theta_0)&=&0\label{equationdF}
\end{eqnarray}
However, it is not possible to extract $\theta_0$ as a function of
$\theta$ and $\tau$ from  Eq.~(\ref{equationdF}), in order to obtain a
closed expression $F(\tau,\theta,\theta_0(\tau,\theta))$ for the
caustics. We will use approximate expressions of
$F$ close to the shocks.

In the neighborhood of the shock, the trajectories can be
approximated by straight lines, with the following expression:
\begin{eqnarray}
\theta(\tau,\theta_0) & = & v_0 \left(\tau-K_0 \right)
\label{theta}
\end{eqnarray}
where  $v_0=v(K_0)$ is the speed of the particle at the
bottom of the potential well.

The envelope of this family of characteristics  can then be
obtained~\cite{Zauderer} by solving the following system of equations
\begin{eqnarray}
\label{eqncaustique}
\theta(\tau,\theta_0) & = & v_0 \left(\tau-K_0 \right) \\
0 & = & v_0^{'}\left(\tau-K_0 \right) -v_0 K^{'}_0
\end{eqnarray}
where the primes denotes the derivative with respect to $\theta_0$.
Using the conservation of energy, we easily
get $v_0=-\sin{\theta_0}$ and we obtain thus
\begin{eqnarray}\label{sysysys}
\tau &=&\tan{\theta_0}\ K^{'}_0+K_0\\
\theta &=&-\displaystyle\frac{\sin^2\theta_0}{\cos\theta_0}\
K^{'}_0\quad .
\end{eqnarray}

It is however possible to go further by using higher order terms
in the expression of $\theta$ given in Eq.~(\ref{theta}), i.e. by
considering curves rather than simply straight lines. We have thus
\begin{eqnarray}\label{thetasuivant}
\theta(\tau,\theta_0) & =& \theta(K_0)
+\frac{d\theta}{d\tau}_{|K_0} \left( \tau-K_0 \right)
+\frac{1}{2}\frac{d^2\theta}{d\tau^2}_{|K_0}
\left( \tau-K_0 \right)^2 \nonumber\\
 &&\hskip 2truecm +\frac{1}{6}\frac{d^3\theta}{d\tau^3}_{|K_0}
\left( \tau-K_0 \right)^3 +\dots\\
& =&  v_0 \left(\tau-K_0 \right)
 -\frac{v_0}{6}\left(\tau-K_0 \right)^3 +\dots\label{replaced}
\end{eqnarray}
where we have replaced in Eq.~(\ref{replaced}) the derivatives by their
expressions. Let
us note in particular, that all derivatives of even order will
vanish because of the parity property of $\theta$. Using the
development of $K_0$ at the same order, we finally end
with the next order approximation for the caustics
\begin{eqnarray}
\tau& = & \tau_s+3a\theta_0^2 +
\left(5b+\frac{2a}{3}\right)\theta_0^4  \nonumber\\
& &+\left(7c+\frac{4b}{3} +\frac{4a}{15}
-\frac{8a^3}{3}\right)\theta_0^6
+O(\theta_0^8)\label{eqpourt}\\
\theta & = & 2a\theta_0^3 + \left(4b+\frac{a}{3}\right)\theta_0^5
\nonumber\\
& & +\left(6c+\frac{2b}{3}+\frac{31a}{180}-4a^3\right) \theta_0^7
+O(\theta_0^9)\label{eqpourtheta}
\end{eqnarray}
where $(a,b,c)=\frac{\pi}{8}\left(1, \frac{11}{48}, \frac{173}{2880}\right)$.

\begin{figure}
{\resizebox{0.5\textwidth}{!}{\includegraphics{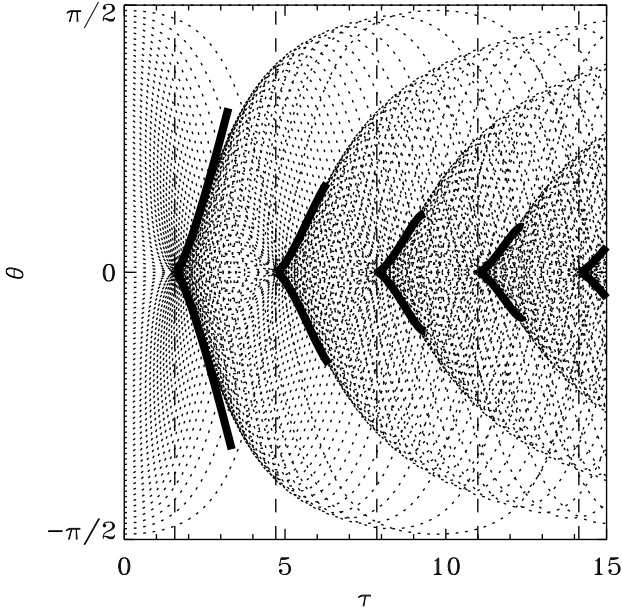}}
}
\caption{Superposition of the caustics over the characteristics of
50 particles evenly distributed between $-\pi/2$ and $\pi/2$ at
$t=0$. The analytical formula for the ``chevrons''
(Eq.~\ref{eqpourt} and Eq.~\ref{eqpourtheta}) is superimposed
(bold curves). The vertical dashed lines shows the appearance time
of the chevrons $t_n=t_s*(2n-1)$.} \label{chevrons2}
\end{figure}
It is possible to see in Fig.~(\ref{chevrons2}) that the agreement
is excellent and that this procedure is really accurate to
describe the first chevron. Moreover, one can check that because
of the non isochronism of the oscillations, the more energetic
particles arrive too late in the bottom of the well to take part
to the creation of the caustics, i.e. the singularity in the
density space.  Consequently, the untrapped particles of the real
dynamics, visible in Fig.~(\ref{carac}b), but absent of the
averaged dynamics of the associated Burgers equation (see
Fig.~\ref{carac}a), are irrelevant and do not affect the dynamics of the
caustic formation.

\begin{figure}
{\resizebox{0.5\textwidth}{!}{\includegraphics{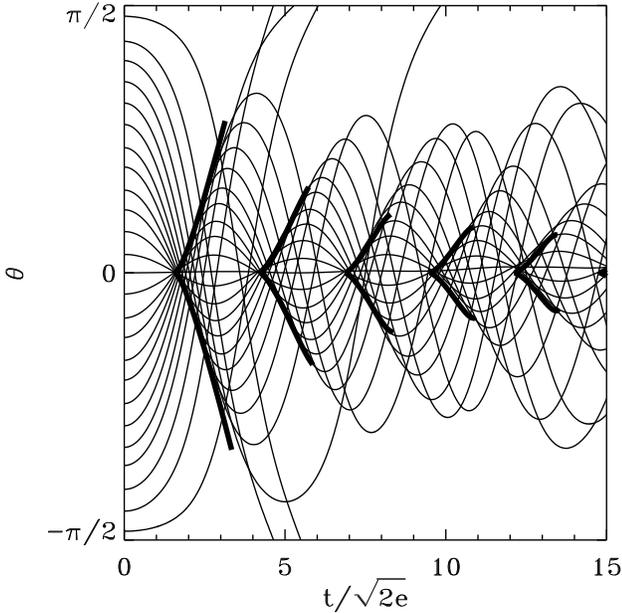}}}
\caption{Superposition of the caustics over the trajectories of
100 particles evenly distributed between 0 and $2\pi$ at $t=0$ for
an energy $e=3\ 10^{-6}$. The time occurrence of the shocks was
reduced by a factor 0.84 (see text).}
\label{trajectoireetcaustique}
\end{figure}

This is confirmed by Fig.~(\ref{trajectoireetcaustique}) where the
caustics determined by the Burgers' approach are superimposed on
the real trajectories: the shape of the chevrons is very well
described. However, the time occurrence of the shock was reduced
by a factor 0.84. The reason, clarified in the next section, is
due to the renormalization of the oscillation frequency $\omega$
which modifies the oscillation period of the particles. As shown
below using a self-consistent procedure,  $\omega$ is reduced with
respect to the plasma frequency~$\omega_p$. As the potential
height is inversely proportional to $\omega$ this leads therefore
to a faster appearance of the shocks.

As already mentioned, the time occurrence of the $n$th shock is
$\tau_n(\theta_0) = (2n-1) \tau_s$ which  means that its shape
 will be obtained simply by replacing in Eq.~(\ref{eqncaustique}) $K_0$ by
$(2n-1)K_0$.
Therefore, the lowest order approximation of the $n$th shocks is
\begin{eqnarray}
\label{chevrons} \theta & = & 2a (2n-1) \left(\frac{\tau-\tau_n}{3a(2n-1)}\right)^{3/2}\\
 \theta & \propto & \frac{\left(
t-(2n-1)t_s \right)^{3/2}}{\sqrt{2n-1}}
\end{eqnarray}
where $n$ is the number of appearance of the "chevron". The
$1/\sqrt n$ factor accounts for the shrinking of the
``che\-vrons'' and the bolded line in Fig.~(\ref{chevrons2})
attests that this expression is particularly accurate.

Similar caustics are encountered in astrophysics, to explain the large
scale structure of the universe: clusters and super clusters of
galaxies are believed to be reminiscent of three dimensional caustics
arising from the evolution of an initially slightly inhomogeneous
plasma~\cite{Vergassola,Zeldovich}.

\subsection{Probability distribution}

\begin{figure}
\resizebox{0.5\textwidth}{!}{
\includegraphics{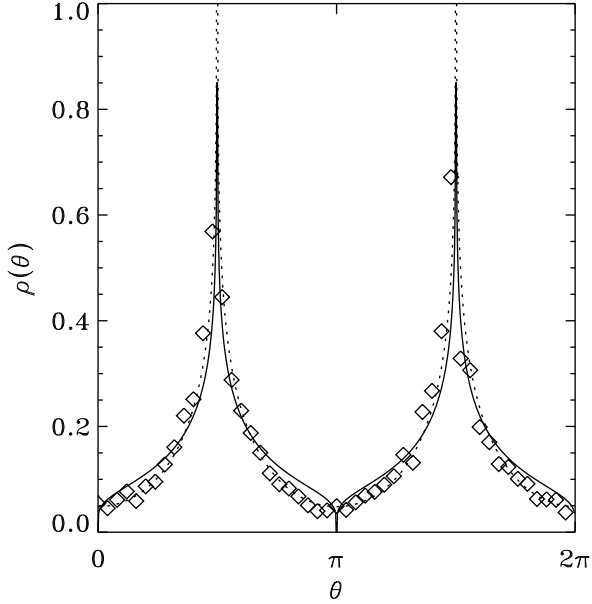}}
 \caption{ Equilibrium distribution. Comparison between the
numerical results (diamonds) and the analytical formula
(\ref{densite}) (solid line)
 in the case $e\simeq10^{-4}$. The dotted line
corresponds to the formula~(\ref{distr}).} \label{histotheor}
\end{figure}

At this point, it would be important to calculate the density
distribution in the vicinity of the singularity. For this hydrodynamical approximation, it would be fully
determined by the initial conditions, i.e. the initial velocity
distribution. Let us recall that we earlier found~\cite{drh}, that
the following analytical formula
\begin{equation}
{\cal P}(\theta) = \frac{1}{2\pi} \left( 1 - \log (2 |\sin \theta| ) \right)
\label{distr}
\end{equation}
was an excellent approximation (see Fig.~\ref{histotheor}), even
if we didn't have any theory for its derivation. Here using the
Burgers' approach, let us obtain a similar result.

Along the particle trajectory, one has
\begin{equation}
\dot \theta=\sqrt{\frac{\cos 2\theta-\cos 2\theta_0}{2}}
\end{equation}
The time $dt(\theta)$ spent by a particle close to a
position $\theta$ is inversely proportional to its velocity $\dot
\theta(\theta_0)$.  Therefore the trajectory parametrized by
$\theta_0$ (and initiated in $\theta_0$) gives a contribution to the
density
\begin{eqnarray}
\rho_{\theta_0}(\theta)d\theta&=&\frac{dt(\theta)d\theta}{\displaystyle
\int_{-\theta_0}^{\theta_0}dt(\theta) }
=\frac{2\sqrt{2}\ d\theta}{T(\theta_0)\sqrt{{\cos 2\theta-\cos 2\theta_0}}}
\end{eqnarray}
where $T(\theta_0)$ is the period of the trajectory. By recalling
that the initial distribution is homogeneous in these numerical
computations, one obtains the probability density by averaging
over the time, the density of characteristics at a fixed position.
We obtain:
\begin{eqnarray}
\label{densite}\rho(\theta)  \propto  \int_{2\theta}^{\pi}{\frac{2\sqrt{2}
\ du}{T(u)\sqrt{\cos2\theta-\cos 2u}}}
\end{eqnarray}
for $\theta=[0,\pi/2]$ whereas
the whole distribution is obtained by symmetry and $\pi$-periodicity.

Due to the numerous approximations, the agreement for long
time is not perfect, but equation (\ref{densite}) gives a good
result as shown by Fig.~(\ref{histotheor}).  The disagreement is
visible in $0$ and $\pi$, i.e. close to the separatrix.
Actually, as we will see in the following, the amplitude of the
effective potential slightly oscillates; this allows the existence
of untrapped particles which smooth the density.

As a conclusion, this hydrodynamical approach has proved to be
very successful, since it explains qualitatively and
quantitatively, for short times, the main features of the
bicluster formation. Nevertheless, the stability
and the long time behavior of the structure is not yet
understood: this is the object of the next section.

\section{Long time evolution of the bicluster}
\label{lagrangianeff}

Since the thermodynamics of the antiferromagnetic HMF model
predicts a perfectly homogeneous equilibrium state, the bicluster
is a non equilibrium structure. However, its long time stability,
as well as the fact that it is reached starting from a variety of
initial conditions suggest some statistical explanation.

In addition, we have learned from the previous section that
particle motion may be split into two parts: a small amplitude
rapid oscillation superimposed to a large amplitude slow motion.
The idea is thus to look for an effective Hamiltonian dynamics
which would only retain the slow motion, and would be appropriate
for a statistical description.

\subsection{The fast variables}

The complete Lagrangian of the antiferromagnetic HMF being
\begin{eqnarray}\label{completelagrangian}
L\left( {\theta_i},{\dot{\theta_i}}\right) & = & \sum_{i=1}^{N}
\frac{\dot{\theta_i}^2}{2}
-\frac{1}{2N}\sum_{i,j}\cos(\theta_i-\theta_j)\quad,
\end{eqnarray}
the equations of motion are obtained by minimizing the action $S=\int Ldt$
with respect to the
functions $\theta_i$. Taking advantage of the two well defined time
scales, we introduce the following ansatz:
\begin{equation}
\theta_i=\theta_i^0(\tau)+\varepsilon f_i(t,\tau)\label{ansatzz}
\end{equation}
where $\tau=\varepsilon t$ and $\varepsilon=\sqrt{2e}$ as already
defined in the section~(\ref{nonlinear}). The $f_i$'s represent
the small rapid oscillations and the $\theta_i^0$ the slow
variables which we are finally interested in. The idea is then to
use a variational approach (least action principle), first to
obtain the equations of motion of the fast variables  $f_i$, and
then to reintroduce the solutions into the action. In a
 second stage, averaging over the fast time $t$, we will obtain a
variational principle for the slow variables.  This approach,
inspired by~\cite{witham}, has the double advantage of allowing to
obtain an Hamiltonian system for the slow variables, and the one
of exhibiting a natural conservation law for this dynamics (the
adiabatic invariant). Appendix~A briefly presents the use of
method~\cite{witham} for a slowly modulated harmonic oscillator,
in order to emphasize its power for a very simple model, away from
the rather technical context presented here.

We first notice that since the energy of the system is by
definition of order $\varepsilon^2$, the sums $1/N \sum \cos\theta_i^0$ and
$1/N \sum
\sin\theta_i^0$, which represent the two components of the magnetization
vector in terms of the slow
variables $\theta_i^0$, are of order $\varepsilon$.  We thus define
\begin{eqnarray}
\varepsilon M_{1x}^0 &=&\frac{1}{N} \sum_i \cos{\left(\theta_i^0+\psi\right)}\\
\varepsilon M_{1y}^0 &=&\frac{1}{N} \sum_i \sin{\left(\theta_i^0+\psi\right)}
\end{eqnarray}
where  the phase $\psi$  $\in [0,\pi]$ is chosen such that
the scalar dynamical indicator of the clustering is
\begin{equation}\label{defM2}
|M_2|=\left|\frac{1}{N}\sum_i\exp\left(2i\theta_i^0\right)\right|=
\frac{1}{N}\sum_i\cos2\left(\theta_i^0+\psi\right).\end{equation}
We now introduce the ansatz~(\ref{ansatzz}) into the
Lagran\-gian~(\ref{completelagrangian}), and develop the cosine up to
order $\varepsilon^2$, obtaining the new Lagrangian~$L_2$ that depends on the $\theta_i^0$'s, the $f_i$'s and
their time derivatives
\begin{eqnarray}
L_2&=&L_2\left({\theta_i^0},{\frac{d\theta_i^0}{d\tau}},{f_i},{\frac{df_i}{dt}}\right) \nonumber \\
&=& \frac{\varepsilon^2}{2} \sum_i \left[\left(\frac{d\theta_i^0}{d\tau}
\right)^2
+2\frac{d\theta_i^0}{d\tau}\frac{df_i}{dt}+\left(\frac{df_i}{dt}\right)^2
\right] \nonumber \\
&-&\frac{\varepsilon^2N}{2}\Bigl[(M^0_{1x})^2+(M_{1y}^0)^2\Bigr]+\frac{\varepsilon}{N}
 \sum_{i,j}f_i\sin{(\theta_i^0-\theta_j^0)}
\nonumber \\
&-&\frac{\varepsilon^2}{2N}\sum_{i,j}f_i f_j \cos(\theta_i^0-\theta_j^0) \quad.
\quad\label{lagranL1}
\end{eqnarray}
From this expression, considering $\tau$ as a constant, we write the equations of motion for the fast variables
$f_i$'s. We get
\begin{eqnarray}
\ddot{f_i}
&=M_{1x}^0\sin{\left(\theta_i^0+\psi\right)}&-M_{1y}^0\cos{\left(\theta_i^0+\psi\right)}\nonumber\\
&&-\frac{1}{N}\sum_{j}f_j\cos\left(\theta_i^0-\theta_j^0\right)\quad.\quad
\label{equationpourF}
\end{eqnarray}
This is a linear second order equation, with constant coefficients
with respect to the fast time, whose solution thus requires the
diagonalization of the $N$x$N$  matrix
\begin{equation}
  \label{eq:matrixT}
  {\mathbf T}=\left[T_{ij}\right]=\frac{1}{N}\left[\cos(\theta_i^0-\theta_j^0)\right]\quad,
\end{equation}
which has only two non zero eigenvalues (see Appendix~B)
\begin{equation}\label{eq:freqprop}\omega_\pm^2=\frac{1\pm|M_2|}{2}\quad.\end{equation}
The eigenvectors corresponding
to $\omega_\pm^2$ are respectively
\begin{eqnarray}
  \label{eq:vecprp}
  {\mathbf X_+}&=&\left[\cos(\theta_i^0+\psi)\right]_{i=1\ldots N}\\
{\mathbf X_-}&=&\left[\sin(\theta_i^0+\psi)\right]_{i=1\ldots N}\quad.
\end{eqnarray}
 The general solution for the
$f_i$'s is therefore
\begin{eqnarray}
f_i(t,\tau)
&=&\left[\sqrt{2} A_+\sin(\omega_+t+\varphi_+)-\frac{M_{1y}^0}{\omega_+^2}\right] \cos(\theta_i^0+\psi)  \nonumber\\
&+&\left[\sqrt{2} A_-\sin(\omega_-t+\varphi_-)+\frac{M_{1x}^0}{\omega_-^2}\right] \sin(\theta_i^0+\psi)\quad.\qquad
\end{eqnarray}

\subsection{The slow variables}

\label{slowvariables}

\begin{figure}
\resizebox{0.5\textwidth}{!}{\includegraphics{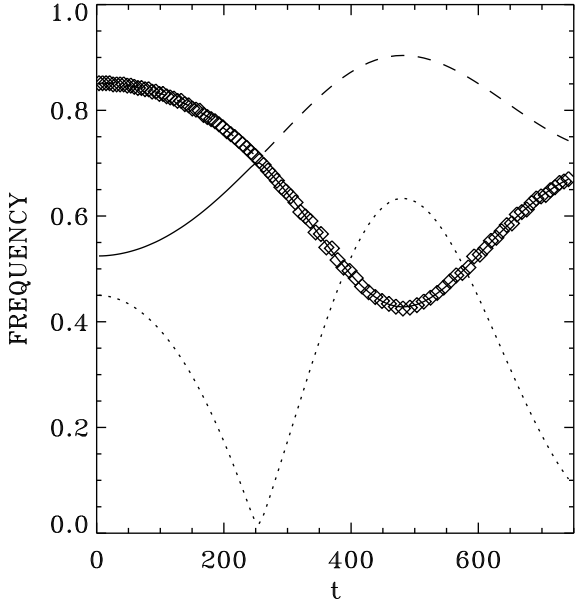} }
\caption{The diamonds presents the evolution versus time of the
main
  frequency for $N=200$ particles of the original
  Hamiltonian~(\ref{hamiltonian}).  We have superimposed not only $\omega_+$
  (dashed line) and $\omega_-$ (solid line) using Eqs.~(\ref{eq:freqprop}),
  but also  the numerically computed $|M_2|$ (dotted line). We
  clearly see the exchange of frequency when  $|M_2|$ touches zero.}
\label{croisementvp}
\end{figure}

The previous solution for the $f_i$ suggests the following new ansatz
\begin{eqnarray}
f_i&=& \left[\sqrt{2} A_+(\tau)\sin(\phi_+(t))+a_+(\tau)\right]\ \cos(\theta_i^0+\psi)\nonumber \\
 &+& \left[\sqrt{2}A_-(\tau)\sin(\phi_-(t))+a_-(\tau)\right]\ \sin(\theta_i^0+\psi)\label{fiansatz}
\end{eqnarray}
where $\phi_\pm(t)$ are fast variables, and $d\phi_\pm/dt$, $A_\pm(\tau)$, and
$a_\pm(\tau)$ are slow variables. We introduce~(\ref{fiansatz}) in the
Lagrangian~(\ref{lagranL1}), and we average over the fast variables
$\phi_\pm$.  Dropping the $\varepsilon^2$ overall factor, the averaged
Lagrangian reads
\begin{eqnarray}
L_{eff} &=& \frac{1}{2}\sum_{i=1}^N \left(\frac{d\theta_i^0}{d\tau}\right)^2
+\frac{N}{2}\left[A_+^2\omega_+^2
\dot \phi_+^2+ A_-^2\omega_-^2\dot \phi_-^2\right]
 \nonumber \\
&&-N\Biggl[\frac{{M_{1x}^0}^2+{M_{1y}^0}^2}{2}+M_{1y}^0 a_+ \ \omega_+^2- M_{1x}^0 a_- \ \omega_-^2   \nonumber \\
&&\qquad+\left({A_+^2}+{a_+^2}\right) \frac{\omega_+^4}{2} +\left(A_-^2+a_-^2\right) \frac{\omega_-^4}{2}\Biggr] \quad.
\end{eqnarray}
Due to the averaging, the variables $\phi_\pm$ are cyclic, $P_\pm$, the conjugate momenta
of $\phi_\pm$, are conserved quantities. Their expression is
\begin{equation}
  \label{eq:pplusmoins}
  P_\pm=\frac{\partial L_{eff}}{\partial\dot \phi_\pm }={NA_\pm^2\omega_\pm^2\dot \phi_\pm}\quad .
\end{equation}
As the Lagrangian does not depend on the time derivatives of
$A_+$, $A_-$, $a_+$, and $a_-$, there is no Legendre transform
on these variables and the Hamiltonian reads
\begin{eqnarray}
H_{eff} &=& P_+\dot\phi_++P_-\dot\phi_-+\sum_{i=1}^Np_i\dot \theta_i^0-L_{eff}\\
&=&\frac{P_+^2}{2NA_+^2 \omega_+^2}
+\frac{P_-^2}{2NA_-^2 \omega_-^2}+\sum_{i=1}^N \frac{{p_i^0}^2}{2}
\nonumber \\
&&+N\Biggl[\frac{{M_{1x}^0}^2+{M_{1y}^0}^2}{2} +M_{1y}^0 a_+ \omega_+^2- M_{1x}^0 a_- \omega_-^2
 \nonumber \\
&&\qquad+\left(A_+^2+a_+^2\right)\frac{\omega_+^4}{2}+\left(A_-^2+a_-^2\right)\frac{\omega_-^4}{2} \Biggr]\quad.
 \label{hamiltonien}
\end{eqnarray}
In the absence of conjugate variables of the amplitudes $A_\pm$, the
corresponding Hamilton's equation are simply given, from the least
action principle, by $\partial_{A_\pm} H_{eff}=0$. 
 Together with the equations
(\ref{eq:pplusmoins}), this leads to the following expressions for the
frequencies \begin{eqnarray}
\frac{d\phi_\pm}{dt} &=& \omega_\pm\quad,  \label{omegax}
\end{eqnarray}
expected of course from the previous study of the matrix~${\mathbf
  T}$, and for the amplitudes
\begin{eqnarray}
A_\pm^2    &=& \frac{P_\pm}{N\omega_\pm^{3}}\label{Ax}\quad.
\end{eqnarray}
Finally, from the equations $\partial_{a_\pm} H_{eff} =0$, we find
\begin{equation}
a_+ =-\frac{M_{1y}^0}{\omega_+^2} \quad\mbox{and}\quad
a_- = \frac{M_{1x}^0}{\omega_-^2} \label{aplusmons}\quad.
\end{equation}
Reintroducing expressions (\ref{Ax}) and (\ref{aplusmons})  in
the Hamiltonian (\ref{hamiltonien}), we end up with the \emph{effective}
Hamiltonian describing the slow motion of the particles
\begin{eqnarray}
H_{eff} & = & \sum_i \frac{{p_i^0}^2}{2}
+P_+\omega_++ P_-\omega_- \label{heff}\quad.
\end{eqnarray}
The evolution of the full system is then approximated by the
dynamics of this effective Hamiltonian, the constant $P_+$ and
$P_-$ being determined by the initial conditions. However, a
difficulty arises when the two eigenvalues crosses, ie when
$|M_2|$ becomes $0$. Since $\psi$ is defined as half the phase of
the complex number $M_{2x}+iM_{2y}$, it experiences at this point
a $\pi/2$ jump, and consequently the eigenvectors $X_\pm$ are
inverted: this phenomenon of eigenvalue crossing is illustrated by
Fig.~\ref{croisementvp}.

\begin{figure}
{\resizebox{0.5\textwidth}{!}{\includegraphics{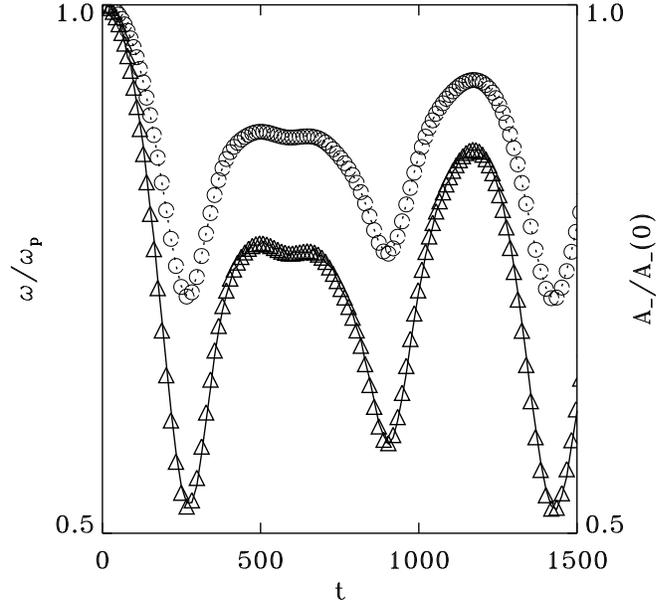}}}
\caption{ Comparison between the numerical (triangles) and
analytical (solid line for $\omega_-$ given by Eq.~(\ref{eq:freqprop})) frequency
of oscillations. Comparison between the numerical (circles) and
analytical (solid line for Eq.~(\ref{Ax})) amplitude $\varepsilon$ of
oscillation .  $e=3.6 10^{-5}$ and $t_{\mbox{\tiny max}}=1500$.}
\label{epsilonperiode}
\end{figure}

However, in most of the numerical experiments conducted in~\cite{drh},
only one fast frequency, $\omega_-$ was excited.  From now on, we will therefore
concentrate on the case $P_+=0$, and will not be concerned by the
phenomenon of eigenvalue crossing.  Dropping the subscript for
$P$, we consider the Hamiltonian
\begin{eqnarray}
H_{eff} & = & \sum_i \frac{{p_i^0}^2}{2}+ P\sqrt{\frac{1-|M_2|}{2}}
\label{heff2}
\end{eqnarray}
The corresponding equations of motion are
\begin{eqnarray}
\ddot{\theta_i} & = & -\frac{P}{N\sqrt{2}\sqrt{1-|M_2|}}\sin{2\left(\theta_i+\psi\right)}
\label{eqofmotion}
\end{eqnarray}
which have to be compared with Eq.~(\ref{burgerslag}): this is still a
pendulum equation, but the potential amplitude, depending now self
consistently on the particles motion through $M_2$, explains the
presence of untrapped particles, shown in Fig.~(\ref{carac}b). In
addition, we have now the right expression for the frequency of the
fast oscillations $\omega=\omega_-$, and a new conserved quantity has been
identified: the adiabatic invariant $P$. Using Eq.~(\ref{Ax}), we
obtain the $A_-(M_2)$ relation which, together with $\omega(M_2)$, is
perfectly verified by numerical simulation as attested by
Fig.~\ref{epsilonperiode}.

The full efficiency of this procedure is that it preserves
completely the Hamiltonian structure of the problem,  making it
 well suited for a statistical treatment, presented
below. However, it is important to emphasize before, that  all
fast variables having disappeared, this procedure induces a huge
gain (of order $1/\varepsilon$) for numerical simulations. This
allows to study accurately the statistics and the dynamics of this
effective Hamiltonian up to extremely long times, giving
conclusion directly relevant to the original model.

\begin{figure}
\resizebox{0.5\textwidth}{!}{\includegraphics{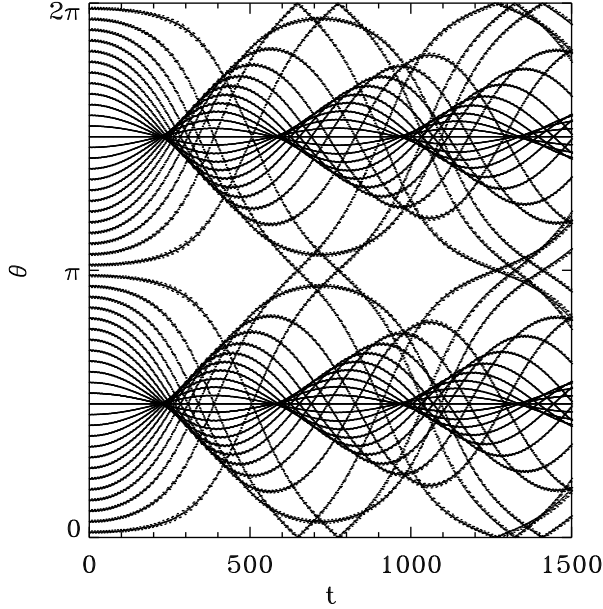}}
\caption{Spacio-temporal evolution of 50 particles. Results given
by the original Hamiltonian~(\ref{hamiltonian}) (resp. effective
Hamiltonian~(\ref{heff})) with (resp. without) the fast small
oscillations superimposed: they are almost indistinguishable.
The initial condition corresponds to particles evenly distributed
on the circle and with a sinuso\"{\i}dally modulated impulsion
corresponding to an energy density $e=2.5\ 10^{-5}$ and
  $\varepsilon=5\ 10^{-3}$.}
\label{compatrajperfect}
\end{figure}

\subsection{Study of the effective Hamiltonian}
\subsubsection{Statistical mechanics of the effective Hamiltonian}
We would like to describe the bicluster as an equilibrium state of the
effective Hamiltonian~(\ref{heff2}).  Fortunately, due to the special
form of the potential, which depends only on the global variable
$|M_2|$, the statistical mechanics of the effective Hamiltonian are
exactly tractable in the microcanonical ensemble.

\begin{figure}
\resizebox{0.5\textwidth}{!}{\includegraphics{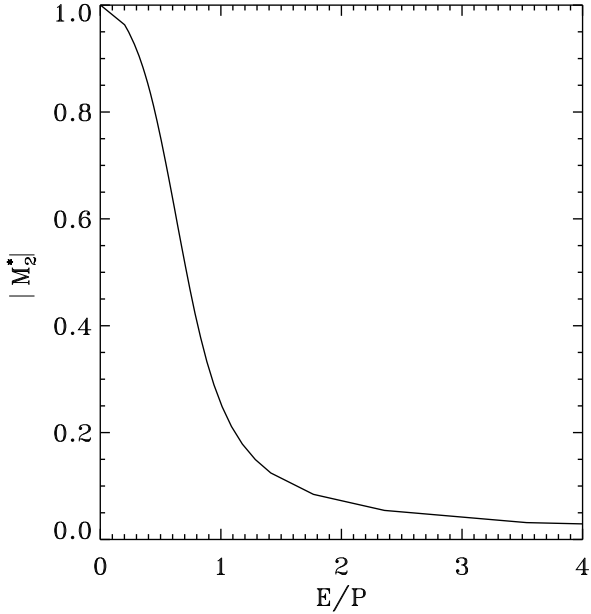}}
\caption{Statistical prediction of the dynamical indicator  $|M_2|$ as a 
function of the ratio between the energy $E$ and the adiabatic invariant $P$.}
\label{solutionthermo}
\end{figure}

There are two conserved quantities: the total energy~$E$ and the
total angular momentum. As the latter only creates a global
rotation of the system totally decoupled from the rest of the
dynamics, we can consider the total energy without
restriction. The density of states $\Omega(E)$ at a given energy,
written as an integral in which the energy is split in two parts,
a kinetic and a potential one, has the following expression
\begin{eqnarray}
\Omega(E) & = & \int dV \: \Omega_{kin}(E-V) \: \Omega_{pot}(V) \nonumber \\
          & \propto & \int d|M_2| \: \Omega_{kin}\left(E-V(|M_2|)\right) \:
          \Omega_{conf}(|M_2|)\qquad
\label{OmegaE}
\end{eqnarray}
where the sign $\propto$ accounts for an unimportant constant.
$\Omega_{kin}$ (resp. $\Omega_{pot}$) is the density of
states corresponding to the kinetic (resp. potential) part of the
Hamiltonian, and $\Omega_{conf}$ is the density of angular
configuration giving rise to a given $|M_2|$: since the potential only depends
on $|M_2|$, $\Omega_{pot}(V)$ is directly proportional to
$\Omega_{conf}(|M_2|)$. Their expressions are
\begin{eqnarray}
\Omega_{kin}(E) & = & \int_{-\infty}^{+\infty}\prod_i dp_i\ \delta \left(
\frac{1}{2}\sum_i p_i^2-E\right) \\
\Omega_{conf}(|M_2|) & = & \int_{0}^{2\pi}\prod_i d\theta_i\ \delta \left(
\frac{1}{N}\left(\sum_i \cos 2\theta_i\right)^2 \right. \nonumber \\
&&+\frac{1}{N}\left.\left(
\sum_i \sin 2\theta_i\right)^2-N|M_2|^2\right) \quad .
\end{eqnarray}
Using the classical result for a perfect gas, we obtain
\begin{eqnarray}
\Omega_{kin}(E) & = & \frac{2\pi^{N/2}}{\Gamma(N/2+1)}\
E^{N/2}\quad .
\end{eqnarray}
To compute $\Omega_{conf}(|M_2|)$, it is possible  to use the inverse
Laplace transform of the Dirac delta function
\begin{equation}
\delta(x) = \int_{\Gamma} dp\: e^{px} \quad, \nonumber
\end{equation}
where $\Gamma$ is a path in the complex plane running from
$\Im(p)=-\infty$ to $\Im(p)=+\infty$. The expression of $\Omega_{conf}$ becomes
\begin{eqnarray}
\Omega_{conf}(|M_2|) & = & \int_{\Gamma} dp \:e^{-Np|M_2|^2} \int_0^{2\pi}
\prod_i d\theta_i \:  e^{NpM_{2}^2}.\quad
\end{eqnarray}
To compute the integral over the angles, we use now the Gaussian
(also called Hubbard-Stratanovich) transform
\begin{eqnarray*}
e^{ Npm^2} &\propto& \int_{-\infty}^{+\infty}du \:
e^{-N \frac{u^2}{4p}+Nmu}\quad.
\end{eqnarray*}
This decouples the integration over the angles $\theta_i$, and we have
\begin{eqnarray}
\Omega_{conf}(|M_2|) & \propto & \int_{\Gamma} dp \:e^{-Np|M_2|^2} \int du\:dv\:
e^{-\frac{N}{4p}(u^2+v^2)} \nonumber \\
&&\int_0^{2\pi}
\prod_i d\theta_i \:  e^{N\sum_i u\cos 2\theta_i +v\sin 2\theta_i} \\
&\propto& \int_{\Gamma} dp \:e^{-Np|M_2|^2} \int_0^{\infty} r dr
e^{-\frac{Nr^2}{4p}} I_0(r)^N\qquad
\end{eqnarray}
where $r$ is the polar radius associated with $u$ and $v$, and $I_0$ is the
modified Bessel function of order $0$. Using the saddle point method on $p$
and $r$, we have finally
\begin{eqnarray}
\Omega_{conf}(|M_2|) & \propto &
e^{-N\left(p^{\ast}|M_2|^2+\frac{{r^{\ast}}^2}{4p^{\ast}}
-\ln{I_0(r^{\ast})}\right)}\\
& = &  e^{-N\left(2p^{\ast}|M_2|^2 -\ln{I_0(r^{\ast})}\right)}
\end{eqnarray}
where $p^{\ast}$ and $r^{\ast}$ are implicitly defined by
\begin{eqnarray}
\frac{r^{\ast}}{2p^{\ast}} & = & \frac {d\:\ln{I_0(p)}}{dp}_{|r^\ast}  \\
\frac{{r^{\ast}}^2}{4{p^{\ast}}^2} & = & |M_2|^2   \quad.
\end {eqnarray}
Using this result, Eq.~(\ref{OmegaE}) may be evaluated once again
by the saddle point method, which leads after some simple algebra
to the equilibrium value $|M_2|^{\ast}$, defined by the following equation
\begin{eqnarray}
r^{\ast}(|M_2|^{\ast}) & = & \frac{1}{4\sqrt{1-|M_2|^\ast}
\left(\displaystyle
\frac{\sqrt{2}E}{P}-\sqrt{1-|M_2|^\ast}\right)}\quad .
\end{eqnarray}
As shown by Fig.~\ref{solutionthermo}, this equation has only one
solution for any ratio $E/P$; this indicates an absence of
phase transition. As this solution is non zero, a biclusterization
will always  take place; nevertheless in the large energy limit the
particles are almost free and $|M_2|^\ast$ goes of course to zero.

Once $|M_2|^{\ast}$ is known, it is easy to complete the description of the
equilibrium state. The temperature $T$ is given by
\begin{eqnarray}
\label{temperature}
T &=& \frac{2}{N}\langle E_c \rangle= \frac{2}{N}\left\langle E-P\sqrt{\frac{1-|M_2|^\ast}{2}}\right\rangle
\end{eqnarray}
the distribution of velocities is a Maxwellian with temperature $T$, and the
distribution of angles has a gibbsian shape $\rho(\theta)\propto
e^{-V(\theta)/T}$ with the potential
\begin{eqnarray}
V(\theta) &=& \frac{P}{N2\sqrt{2}\sqrt{1-|M_2|^\ast}}\left(1-\cos(2\theta+2\psi)
\right)\label{gibbsian}
\end{eqnarray}
This potential is inferred from the equations of motion~(\ref{eqofmotion}).\\

\subsubsection{Relaxation to equilibrium}

We have therefore now a complete description of the statistical
equilibrium states of the effective Hamiltonian, governed by long
range interactions. It has been noticed by various authors that
the dynamics of such systems may sustain long lived metastable
states before relaxing to equilibrium
\cite{Antoni,AnteneodoTsallis,Latora1,gouda}. Before comparing the
"effective equilibrium" with the structure created by the dynamics
of the real Hamiltonian, it is thus necessary to study the
relaxation to equilibrium. As the relaxation properties of long
range interacting systems is in itself an important problem, we
will consider them now. We will in addition illustrate which
statistical properties of the equilibrium distribution is expected
to be observed in the real dynamics, for large $N$ and on time scale
reasonable for numerical computations.

\begin{figure}
\resizebox{0.5\textwidth}{!}{\includegraphics{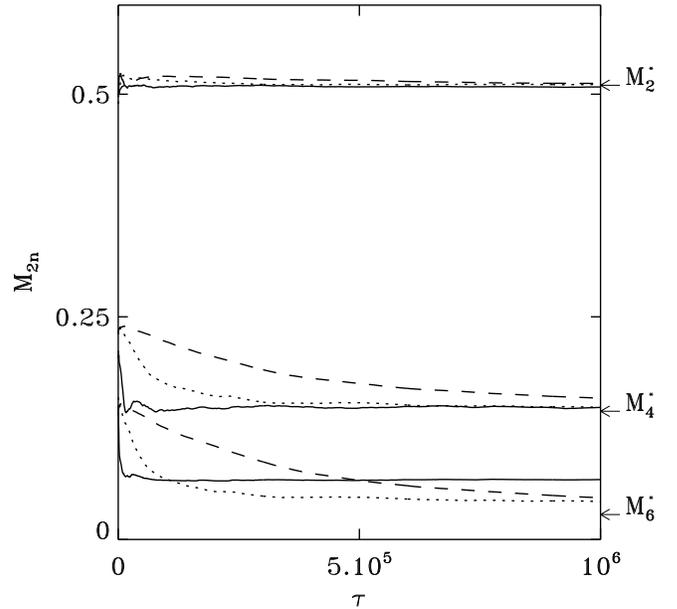}}
\caption{ Relaxation of the first three even  moments $<M_{2n}>$ (time averaged)
toward equilibrium according to  numerical simulations of the
effective Hamiltonian~(\ref{heff2}). The  solid (resp. dotted and dashed)
lines corresponds to results for $N=200$ (resp. $N=800$ and $N=3200$)
particles.  $<M_2>$, $<M_4>$  and $<M_6>$ are represented from top to bottom
($<M_{2n}>$ is a decreasing function of $n$). They converge
toward the equilibrium values $M_2^\star =0.510$, $M_4^\star=0.144$ and
$M_6^\star=0.028$.} \label{relaxation}
\end{figure}

Fig.~\ref{relaxation} illustrates the approach to equilibrium of the
effective Hamiltonian, for three different particle numbers, with initially
immobile and homogeneously distributed particles (this
corresponds to the typical initial condition used in the real
Hamiltonian in \cite{Antoni,drh} for instance). We have represented
$<M_{2n}>(\tau)$, the time averages of $M_{2n}$ from initial time $0$ to time
$\tau$, for $n$ equal to 2, 4 and 6.
 Temporal fluctuations are thus not visible.
We first observe finite size effects concerning the equilibrium values : for
instance  $M_6$ converges for the small system $N=200$ to a value
larger than $M_6^\star$, the equilibrium value.

More interestingly, we observe also that whereas $|M_2|$ quickly reaches its
equilibrium value, the relaxation time
of the successive moments $|M_4|$, $|M_6|$ strongly depends on the system size,
and presumably diverges with $N$. The actual dependence of the relaxation time
of each moment, on the particle number, will not be studied in this paper. It
would for instance address the issue of whether non equilibrium distributions
may be observed in the thermodynamic limit. Such phenomena have indeed
already been observed in other long range interacting
systems~\cite{Latora1,Latora2}.

We conclude that for a large, but finite, particle number, the
moments of the distribution converge toward their equilibrium values.
When $N$ increases, the relaxation time for the high order moments
may however be larger than computationally achievable times and, we thus expect the simulation to exhibit very
slowly evolving non equilibrium structures of the effective
Hamiltonian. Authors of \cite{drh} have indeed found a
distribution of particles different from the Gibbsian shaped
distribution~(\ref{gibbsian}) predicted by the equilibrium
thermodynamics of the effective Hamiltonian (see
Fig.~\ref{histotheor}).

On the contrary our analysis shows that $|M_2|$ quickly reaches
its equilibrium value $|M_2|^\ast$. For this reason, given the
computational time achievable, in the next section, we only use
this statistical equilibrium indicator for the study of the
structure. This will lead to the correct prediction of the
equilibrium repartition  of potential and kinetic energy, and of
its dependance from initial condition.

\subsection{Comparison with the full Hamiltonian}

If both time scales are clearly separated, Fig.~\ref{compatrajperfect}
emphasizes the striking
agreement between the effective and the real dynamics, for short times. We
will show in this section that the effective
Hamiltonian also provides a good description for the long time
dynamics. 

\begin{figure}
\resizebox{0.5\textwidth}{!}{\includegraphics{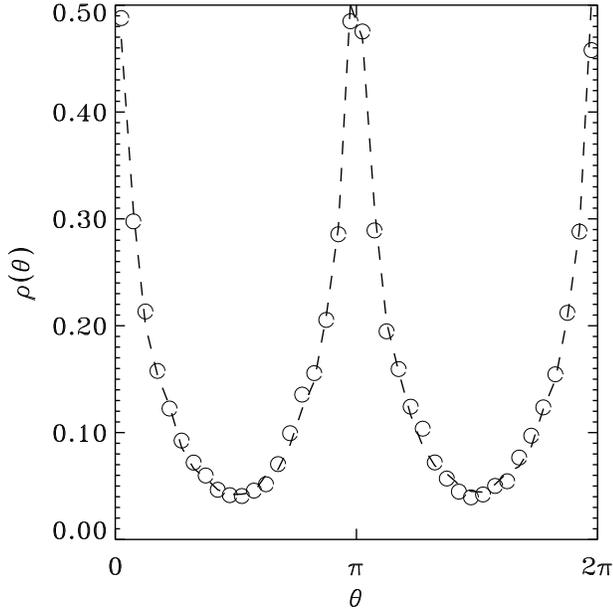}}
\caption{Comparison of the angular density distribution $\rho(\theta)$ of particles obtained 
with the original Hamiltonian~(\ref{hamiltonian}) (represented with circles)
and with the effective one~(\ref{heff2}) (represented with the dashed line).
Both results have been obtained for $N=10^3$ particles and are averaged on 
intermediate times, corresponding to $\tau=10^3 \to 10^4$.
The energy of the original Hamiltonian is  $e=2.5\ 10^{-5}$.} 
\label{compahisto}
\end{figure}

For this purpose, we will first compare the evolution of the non-equilibrium
angular density distribution $\rho(\theta)$ for both dynamics. The results are
reported on Fig.~\ref{compahisto} were circles show the repartition
given by the original Hamiltonian~(\ref{hamiltonian}) whereas the
dashed line shows  the results of the effective dynamics~(\ref{heff2}).
This picture clearly emphasizes that the non equilibrium character of
the dynamics is fully described by the effective dynamics, 
and we can easily conclude that the effective 
Hamiltonian reflects very well  the dynamics
of the real Hamiltonian.

As discussed
in the previous section, according to the computational times used, only the
second momentum of the distribution $M_2$ should correspond to the
equilibrium one. Let us therefore
 analyze the dependence of the equilibrium value of  $M_2$
on the initial conditions. The typically used initial condition for
the numerical simulations in the original system~(\ref{hamiltonian})
is an initially quasi homogeneous distribution of particles with zero
velocity~\cite{drh}. It corresponds to the  ratio  $e/\sqrt{2}P=1$ for
the effective Hamiltonian. According to Fig.~\ref{solutionthermo} ,
this implies an equilibrium value $|M_2|^\ast \simeq0.51$ in perfect 
agreement with the numerical result reported earlier~\cite{drh}.

From these results, 
it is also possible to explain the caloric curve $T\simeq 1.3\ e$,
 reported earlier~\cite{drh}. The total
energy of the full system is divided in three parts: the potential
energy of the small oscillations, the kinetic energy of the small
oscillations and the kinetic energy of the slow movements. The two
first parts are equal in average and form the potential part of
the effective Hamiltonian, whereas the last one is the kinetic
energy of the effective Hamiltonian. Using this remark together
with the values of $e/P$ and $|M_2|^\ast$ at equilibrium, it is easy to
derive the temperature/energy relationship, using (\ref{temperature}). In the case
investigated in \cite{drh}, we find precisely $T=1.3\ e$.

We thus conclude that the dynamics of the effective Hamiltonian
parametrizes very well the dynamics of the real Hamiltonian, for
short as well as for long time. This allows us to predict
statistical properties of the initial system, as for instance the
asymptotic value of  $M_2$ or the partitioning between kinetic and
potential energy. Moreover, let us recall that  the effective
Hamiltonian gives the opportunity to study numerically the
relaxation towards equilibrium of the bicluster, whereas it was
not possible in the real dynamics, because of computational
limitations (let us note that the ratio of the typical time scale
of the two dynamics is of order 100 or larger).

All these findings are great successes of this approach.
Let us nevertheless comment some points that we have not yet addressed.

The first one concerns the limit of validity of our multiple time
scale analysis. It should be noticed that not all initial
conditions with small energy would lead to the formation of the
bicluster. From our analysis, we can conclude that the class of
initial conditions leading to this formation is the one compatible
with the ansatz used: however a precise description of this class
is not known. In particular, we have not study the threshold value
of $\varepsilon$ beyond which such a description breaks down.
For $\varepsilon$ going toward one, the ansatz (\ref{ansatzz})
will first lead to a nonlinear evolution of the slow variable
$f_i$ in place of the linear one given by Eq.~(\ref{equationpourF}). We think that
the critical value $\varepsilon_c$ above which this nonlinear
dynamics do not have any more periodic solution should correspond
to the critical value above which the bicluster can not form,
explaining the transition observed in \cite{drh}. Such an
hypothesis should be tested.

A second point is the phenomenon of level crossing discussed in
section \ref{slowvariables}, leading to an exchange of excitation
of the two modes of the system. This phenomenon is very similar to
level crossing in quantum mechanics adiabatic problems. It
represents a resonance, local in time, in which an interaction
between the various modes may occur, leading to a modification of
the adiabatic invariant. For sake of simplicity, we have
concentrated our work on the case of excitation of only the
smallest frequency, so that such level crossing could not occur. A
more general study would however be of interest.

The last point concerns the validity of our approach for infinite
times.  As noted in the introduction, a recent paper~\cite{FFR}
has shown a long-time degradation of the bicluster, for a very
small number of particles, suggesting its transient
non-equilibrium nature in that case. For a reasonable number of
particles, our numerical results show that such a degradation does
not exist, for computationally achievable times. This point is
thus not addressable numerically. From a theoretical point of
view, results on much more simple systems, show that adiabatic
invariants as described here, should be conserved for a very long
time (exponential when $\varepsilon$ goes to 0). Moreover, such very
long time stability results, would not give any hint on what
happens for larger times (stability or instability).

\section{Conclusion}

\label{Sec-Conclusions}

The surprising formation and stabilization~\cite{drh} of the
bicluster in the HMF dynamics, in contradiction with statistical
mechanics prediction, is now understood: the small collective
oscillations of the bulk of particles create an effective
double-well potential, in which the particles evolve.

On a first stage the dynamics can be described in the context of a
forced Burgers' equation. The particles trajectories caustics
form infinite density regions explaining an initially diverging
density. Early times dynamics is therefore very similar to the
structure formation in Eulerian coordinates for a one dimensional
self-gravitating system~\cite{noullez}. As in this case, because
of the confining potential, the particles can not move apart as
easily after the first caustic has been formed, as in the case of
the free motion on the plane~\cite{noullez}. 

On  much
larger timescales, in order to parametrize the fast oscillations of the
bulk particles, we have performed a variational multiscale
analysis. We then have obtained an Hamiltonian effective model
describing the particle slow motion. This description is in very
good agreement with the initial dynamics. The effective dynamics
allows numerical simulations on time scale much larger than the
initial dynamics, as the rapid oscillations have been filtered
out. We have performed the statistical mechanics of this new
system. The results thus give a statistical explanation of the
bicluster formation and stabilization. The equilibrium properties
can then explain the mean value of the second moment of the
distribution and the repartition between potential and kinetic
energy. Even if this analysis has only been sketched in this work,
the effective dynamics is also a powerful tool to study the very
slow relaxation process versus the real equilibrium structure.

Many physical systems share the main properties of this HMF
dynamics : very fast oscillations self interacting with a slower
motion. In addition to the plasma problem already cited
and always in the context of long range interacting systems, we may
for instance cite the problem of interaction of fast inertia
gravity waves with the vortical motion, for the rotating Shallow
Water or the primitive equation dynamics, in the limit of a small
Rossby Number~\cite{embidmajda}. The main interest of our study is
to provide a toy model in which such a complex dynamics can be
treated and analyzed extensively using powerful theoretical tools.
We point out in particular the usefulness of a variational
approach in the procedure of averaging the rapid oscillations.
This toy model also permits a clear view of the usual problems of
such dynamics\\
(i) Resonances (or level crossing) in   averaging procedures\\
(ii) Unusual relaxation processes due to the long range nature of the
interactions.\\
 The HMF model is to our knowledge the simplest
one in which such phenomena occur.   The study of these
phenomena is thus a natural extension of our work and should be of
interest in analyzing more complex systems.

\acknowledgement We would like to thank  M-C. Firpo, P.
Holdsworth, S. Lepri, J. Lebowitz, F. Leyvraz, A. Noullez, for very helpful
advices. This work has been partially supported by EU contract No.
HPRN-CT-1999-00163 (LOCNET network), the French Minist{\`e}re de la
Recherche grant ACI jeune chercheur-2001 N$^\circ$ 21-31, and the R{\'e}gion
Rh{\^o}ne-Alpes for the fellowship N$^\circ$ 01-009261-01.
This work is also part of the contract COFIN00 on {\it Chaos and
localization in classical and quantum mechanics}.

\section*{Appendix A: Variationnal description of a simple adiabatic dynamics}
\label{appendio}
 Let us consider a slowly modulated harmonic
oscillator with Lagrangian
\begin{equation}\label{lagran}
  L(\dot\theta,\theta,t,\tau)=\frac{\dot\theta^2}{2}-\omega^2(\tau)\frac{\theta^2}{2}
\end{equation}
where $\tau=\varepsilon t$.
We  consider the following ansatz
\begin{equation}\label{ansatzappendix}
\theta(t,\tau)=A(\tau)\sin\varphi(t)
\end{equation}
with  $\dot\varphi=O(1)$ and $\dot A= \varepsilon dA/d\tau$, but,
contrary to the usual asymptotic expansion on the equation of
motion, we will consider the variational approach proposed by
Witham~\cite{witham}. The idea is to separate the two different
time scales of the motion at the level of the action. The average
of the fast variable $t$ yields the following effective
Lagrangian:
\begin{equation}\label{heffappandix}
  {\cal L}=\langle
  L\rangle_t=\frac{A^2}{4}\left(\dot\varphi^2-\omega^2\right)\quad
  .
\end{equation}
It is thus straightforward to obtain the equation of motion of the
effective Lagrangian ${\cal L}(A,\dot A,\varphi,\dot\varphi)$.
They read
\begin{eqnarray}
\frac{d}{dt}\left(\frac{\partial {\cal L}}{\partial \dot
A}\right)&=&\frac{\partial {\cal L}}{\partial A}\\
\frac{d}{dt}\left(\frac{\partial {\cal L}}{\partial \dot
\varphi}\right)&=&\frac{\partial {\cal L}}{\partial \varphi}
\end{eqnarray}
i.e.
\begin{eqnarray}
0=A\left(\dot\varphi^2-\omega^2\right)&\quad\Rightarrow\quad& \dot \varphi=\omega\\
\frac{d}{dt}\left(\frac{A^2\dot\varphi}{2}\right)=0&\quad
\Rightarrow\quad& \dot A^2\dot\varphi=\mbox{cste}
\end{eqnarray}
This method emphasizes a new constant of the motion
$A^2\dot\varphi$, called adiabatic invariant, that is not
exhibited by the usual asymptotic expansion on the equations of
motion corresponding to the original
Lagrangian~(\ref{heffappandix}). This may be of great interest in
complex systems. Moreover, this invariant necessarily appears as
the angle $\phi$ associated to the fast variable $t$ is always
cyclic after averaging. The same reasoning also apply for larger
order ansatzs, showing that an adiabatic invariant should exist at
any order in $\varepsilon$.

In addition, this method preserves the Hamiltonian character of
the problem allowing for instance a statistical mechanics
treatment as presented in this paper.

\section*{Appendix B: Eigenvalues and eigenvectors of the matrix $\mathbf{T}$}
To solve the linear second order equation (\ref{equationpourF}),
we need to diagonalize the  $N$th-order circulant matrix $\mathbf{T}$ defined
by $T_{ij}=\cos(\theta_i-\theta_j)/N$. We introduce the two
vectors $\mathbf{X}(\psi)$ and $\mathbf{Y}(\psi)$, with
coordinates $X_i(\psi)=\cos(\theta_i+\psi)$ and
$Y_i(\psi)=\sin(\theta_i+\psi)$, and $\psi$ an arbitrary phase.
$\mathbf{T}$ is therefore the sum of the projections along these
vectors
\begin{equation}
\mathbf{T} = \frac{1}{N} \left(
\mathbf{X}(\psi)^t\mathbf{X}(\psi)\:
+\:\mathbf{Y}(\psi)^t\mathbf{Y}(\psi)\right)\quad.
\end{equation}
This proves that the image of $\mathbf{T}$ is of dimension two,
and that only two eigenvalues $\lambda_{\pm}$ are nonzero. Let us
restrict now to the two dimensional problem in the
$Vect(\mathbf{X},\mathbf{Y})$ plane (this plane does not depend on
$\psi$ of course), and let us  choose $\psi$ such that $X$ and $Y$
are orthogonal, i.e.
\begin{equation}
(\mathbf{X}.\mathbf{Y})=\sum_i
\cos(\theta_i+\psi)\sin(\theta_i+\psi) = 0\quad.
\end{equation}
This definition of $\psi$ leads to
\begin{equation} |M_2| =
\frac{1}{N}\sum_i \cos(2\theta_i+2\psi)\quad.
\end{equation}
Once we have the two eigenvectors $\mathbf{X_+}=\mathbf{X}(\psi)$
and $\mathbf{X_-}=\mathbf{Y}(\psi)$,  their
 associated eigenvalues  are defined by the following relationship, valid $\forall i$:
 \begin{equation}
   \label{eq:demomega}
   \sum_j\frac{1}{N}\cos(\theta_i-\theta_j)X_{\pm i} =\lambda_\pm X_{\pm i} .
 \end{equation}
leading directly to
\begin{equation}
\lambda_{\pm} = \frac{1\pm|M_2|}{2}\quad.
\end{equation}

\end{document}